\documentclass[conference]{IEEEtran}
\IEEEoverridecommandlockouts
\usepackage{cite}
\usepackage{amsmath,amssymb,amsfonts}
\usepackage{algorithmic}
\usepackage{graphicx}
\usepackage{textcomp}
\usepackage{xcolor}

\usepackage{pgfplots}
\pgfplotsset{compat=1.5}
\usepackage{pgfplotstable}
\usepgfplotslibrary{statistics}
\pgfplotsset{grid style={dotted,gray}}
\usepackage{tikz}
\definecolor{bblue}{HTML}{4F81BD}
\usepackage{soul}

\newcommand{\MYfooter}{\smash{\scriptsize
\hfil\parbox[t][\height][t]{\textwidth}{\centering
\copyright 2021 IEEE. Personal use of this material is permitted. Permission from IEEE must be obtained for all other uses, including reprinting/republishing this material for advertising or promotional purposes, collecting new collected works for resale or redistribution to servers or lists, or reuse of any copyrighted component of this work in other works. DOI: 10.1109/ISWCS49558.2021.9562172}\hfil\hbox{}}}

\makeatletter

\def\ps@IEEEtitlepagestyle{%
\def\@oddfoot{\MYfooter}%
\def\@evenfoot{\MYfooter}}

\makeatother

\DeclareMathOperator{\E}{\mathbb{E}}
\def\BibTeX{{\rm B\kern-.05em{\sc i\kern-.025em b}\kern-.08em
    T\kern-.1667em\lower.7ex\hbox{E}\kern-.125emX}}
\begin{document}


\title{Reliability Analysis of Slotted Aloha with Capture for an OWC-based IoT system}

\makeatletter
\newcommand{\linebreakand}{%
  \end{@IEEEauthorhalign}
  \hfill\mbox{}\par
  \mbox{}\hfill\begin{@IEEEauthorhalign}
}
\makeatother

\author{\IEEEauthorblockN{Milica Petkovic, Tijana Devaja,\\Dejan Vukobratovic}
\IEEEauthorblockA{\textit{University   of   Novi   Sad} \\
\textit{Faculty of Technical Sciences}\\
Novi Sad, Serbia \\
\{milica.petkovic, tijana.devaja\\
dejanv\}@uns.ac.rs}
\and
\IEEEauthorblockN{Francisco J. Escribano}
\IEEEauthorblockA{\textit{Universidad de Alcal\'{a}} \\
\textit{Signal Theory and} \\
\textit{Communications Department}\\
Alcal\'{a} de Henares, Spain \\
francisco.escribano@uah.es}
\and
\IEEEauthorblockN{\v Cedomir Stefanovi\' c}
\IEEEauthorblockA{\textit{Aalborg University} \\
\textit{Department of Electronic Systems}\\
Aalborg, Denmark \\
cs@es.aau.dk}
}

\maketitle

\begin{abstract}

In this article, we consider a random access scheme for an indoor Internet of Things (IoT) framework that uses optical wireless communication (OWC). We focus on a Slotted ALOHA (SA)-based solution where a number of OWC IoT users contend to send data to a central OWC receiver. In any given slot, and for a randomly selected active user, we consider the reliability of decoding the user's data packet at the receiver. This is done by deriving the signal-to-noise-and-interference-ratio (SINR) statistics from a randomly chosen user and evaluating the probability that the user's SINR is below a given threshold. By placing our analysis in the context of an indoor OWC IoT uplink setup, and employing the standard OWC channel model, we investigate the trade-offs between the reliability and the OWC system parameters such as the cell area or the transmitter's semi-angle. We obtain valuable insights into the design of an SA-based random access solution for a typical indoor OWC cell.   
\end{abstract}


\section{Introduction}


The current wave of digital transformation is fueled by increased demands for wireless connectivity, most notably through adoption of the Internet of Things (IoT) paradigm in modern smart infrastructures. 
IoT communications are typically characterized by sporadic and unpredictable device activity involving short data exchanges.
A cost-effective approach to support such communication patterns over a shared wireless medium is to use a random access (RA) protocol, which allows for a more efficient use of the time-frequency resources~\cite{mag, from5to6, modernRA}.
Slotted ALOHA (SA)~\cite{SA, Roberts} is a classical RA scheme, serving as the basis for wireless access solutions implemented in most of today's commercial systems~\cite{lora,sigfox,3gpp}.

Optical wireless communication (OWC) represents a promising future technology to augment classical radio-frequency (RF)-based communications partly addressing the ongoing explosive increase in the demand for additional wireless capacity.
Due to several desirable properties, such as wide and license-free spectrum, high data-rates, low cost and easy deployment, OWC is an appropriate candidate to provide adequate support for the demands of 5G networks~\cite{OWC_MATLAB, b10}. 
OWC can be seen as a promising alternative to RF for IoT technologies, especially in RF-restricted areas where the light-emitting diodes (LED) can be safely used as wireless transmitters.

OWC-based indoor IoT systems have been analyzed in recent literature~\cite{vlc1,vlc2,vlc3}.
Multiple access protocols have been considered for OWC-based IoT systems to ensure throughput efficiency~\cite{MA2,MA3,MA4}. Flexible and throughput-efficient uplink RA mechanisms based on the SA approach, which accommodate sporadic and varying device activity, have been adopted for several OWC-based IoT systems in~\cite{SA1,SA2,SA3}.
In this respect, \cite{SA1,SA2} analyzed the throughput performance of OWC-based IoT  systems considering multi-packet reception and successive interference cancellation.
In~\cite{SA3}, the throughput performance of a two-tier SA multiple-relay system was considered, where the first tier of the uplink between the IoT devices and the relays is performed using indoor OWC, while the second tier assumes an outdoor long-range RF transmission.

In this paper, we present a reliability analysis of one-shot transmission (i.e., the probability that a single transmission attempt will be successful) from the perspective of a randomly selected active user within an OWC-based IoT system. We consider an SA-based uplink communication between a number of IoT devices randomly deployed in an indoor circular area and a single OWC access point (AP). The point of departure of our analysis is the characterization of the optical gain of the channels between the users' and the AP.
The differences among particular optical gains result in an imbalance among the received power values from each user, potentially enabling the capture effect, and consequentially, determining the performance of the access protocol. Specifically, we provide a detailed study of the signal-to-interference-and-noise (SINR) in the system, and use it to investigate the reception reliability 
of a randomly activated user by taking into account the contribution of the interference from other contending users.
The derived expressions (validated through simulation) are used to characterize the trade-offs between the reliability and the OWC system setup parameters. 
We show that the factors determining the geometry of the framework (area where the transmitters are located, distance from the OWC AP, transmitter's semi-angle) have a substantial impact. 
The presented results can be used to asses and optimize the performance of the access protocol.
Specifically, in this paper we evaluate the reliability of SA for a simple case of Bernoulli arrivals and show that there is a non negligible probability that an active user could be decoded in collision slots, effectively lowering the outage probability, i.e., increasing the reliability.  

The rest of the text is organized as follows.
Section II presents the system model. In Section III, the SINR statistics for a random active user is derived. The reliability analysis, constituting the central part of the paper, together with the numerical results, are provided in Section IV. Section V concludes the paper.

\section{System Model for SA-based Indoor OWC IoT}

We consider an uplink communication scenario where a total of $U$ IoT devices contend to access the common OWC AP. The devices are randomly placed on a horizontal plane (a floor or a working plane), while the OWC AP is located at the ceiling (see Fig.~\ref{Fig1}). The SA protocol~\cite{Roberts} is considered for the uplink transmission. 
Every IoT device is active with probability $p_a$ in every slot, independently of the activity of other devices in the same slot.
If a user is active in a slot, it transmits a fixed-length packet fitting the slot.
The number of active users in a given slot is denoted by $U_a$ and assumed to be a fixed (but unknown) parameter used in the analysis.
In Section~IV, we assume a scenario with Bernoulli arrivals in order to model the distribution of $U_a$.

Active devices employ LED-based sources operating in the IR spectrum that use intensity-modulation (IM) binary-format signaling (e.g., non-return-to-zero on-off keying scheme). The OWC photo-detector (PD) receiver implements direct detection (DD) of the light intensity~\cite{OWC_MATLAB}. 

\begin{figure}[b!]
\centerline{\includegraphics[width=0.97\columnwidth]{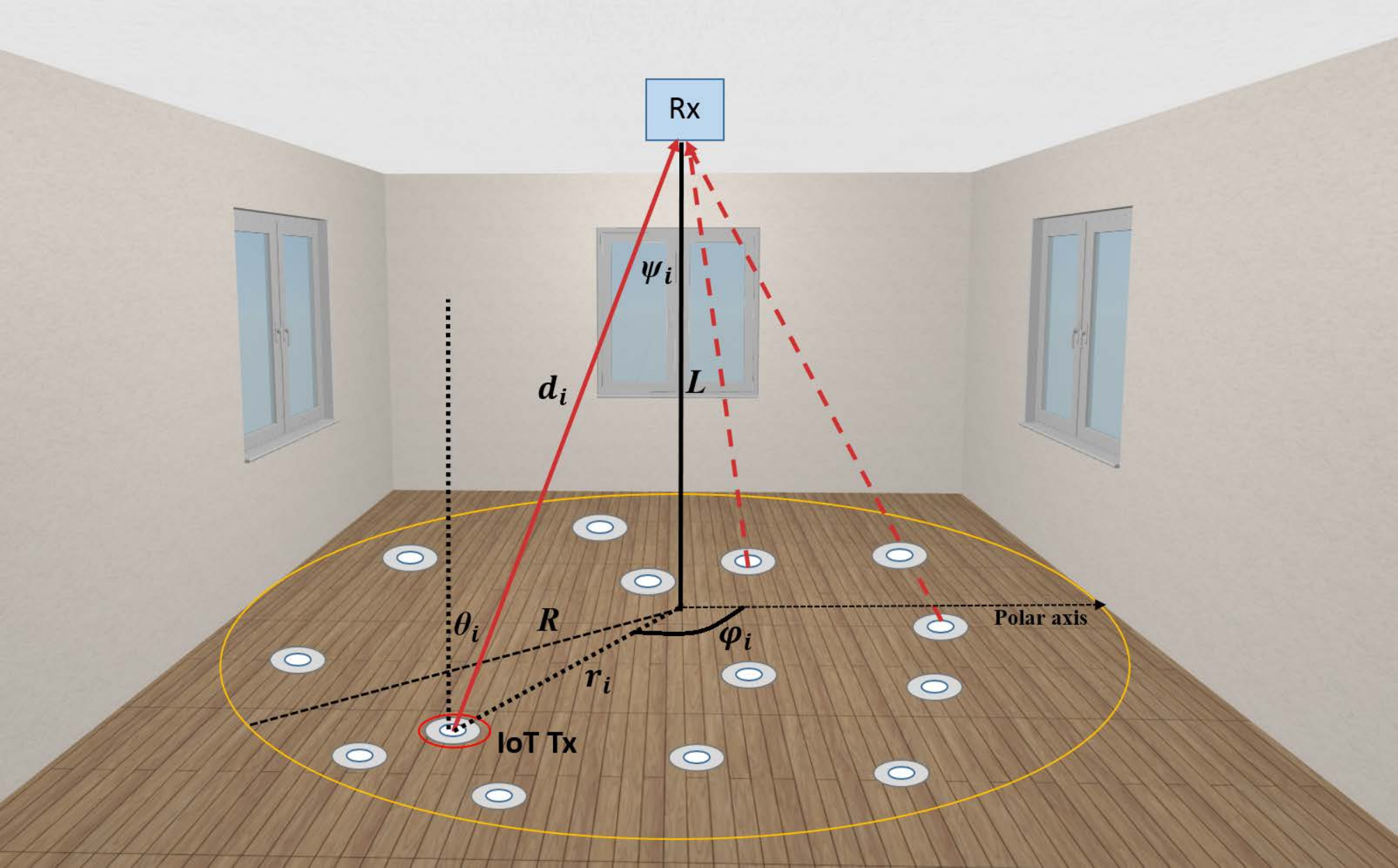}}
\caption{OWC-based IoT system model.}
\label{Fig1}
\end{figure}

Considering any given slot with $U_a$ active users, the light intensity impinging the OWC PD receiver consist of the active users' contribution and the background radiation noise. Assuming that the PD is working in a linear regime (which is a standard and an adequate approximation), the received signal after the conversion to the electrical domain is
\begin{equation}
\begin{split}
y(t)= \sum_{i=1}^{U_a} P_t \eta h_i x_i(t) + n(t)
\end{split}
\label{y1}
\end{equation}
where $x_i(t)$, $i=1, \cdots, U_a$, is the signal sent by the $i$-th user, $P_t$ is the transmitted optical power and $\eta$ the optical-to-electrical conversion coefficient. Further, $h_i \geq 0$ is the optical gain from the $i$-th user to the AP, while $n(t)$ is the Gaussian noise comprising background radiation and thermal noise. In out setup, $n(t)$ is modeled as a zero-mean Gaussian random variable (RV) with variance $\sigma_n^2=N_0B$, where $N_0$ is the noise spectral density and $B$ is the system bandwidth. 

In the system under consideration, if a slot contains transmissions from two or more users, i.e., if $U_a \geq 2$, a packet collision occurs. In that case (in contrast to the classical SA collision channel model), we assume that the OWC AP receiver will attempt to decode the transmission. We are interested in the probability that the receiver will succeed in decoding a randomly selected user among the set of active ones.
Without loss of generality, we will assume that the index of that user to corresponds to $i = U_a$, and henceforth we will refer to that user as the reference user.
For this setup, the received signal in (\ref{y1}) can be rewritten as
\begin{equation}
\begin{split}
y(t)= P_t \eta h_{U_a} x_{U_a}(t) + \sum_{i=1}^{U_a-1} P_t \eta h_i x_i(t) + n(t)
\end{split}
\label{y2}
\end{equation}
where $x_{U_a}(t)$ is the signal sent by the reference user, $h_{U_a} \geq 0$ is the optical gain from the reference user to the PD receiver, and the summation term represents the interference contribution from all other active users except the reference one.

Based on \eqref{y2}, the instantaneous SINR for the active user can be computed as
\begin{equation}
\begin{split}
{\rm SINR} & = \frac{ P_t^2 \eta^2 h_{U_a}^2 }{ \sum_{i=1}^{U_a-1}P_t^2 \eta^2 h_i^2 + \sigma _n^2} \\
& = \frac{\gamma_{U_a}  }{ \sum_{i=1}^{U_a-1} \gamma_i + 1} = \frac{\gamma_{U_a}  }{ \gamma_{\rm I} + 1}
\label{sinr}
\end{split}
\end{equation}
where 
\begin{equation}
\gamma_{U_a}  = \frac{ P_t^2 \eta^2   h_{U_a}^2 }{\sigma _n^2},~\gamma_i  = \frac{ P_t^2 \eta^2 h_i^2}{\sigma _n^2},~\gamma_{\rm I}  =\!\! \sum_{i=1}^{U_a-1} \!\!\gamma_i.
\label{gUa}
\end{equation}
Next, we derive the statistics for an SINR observed by a randomly selected active user.

\section{SINR Statistics for SA-based OWC IoT}
 

We assume that the devices are randomly and uniformly distributed over the indoor coverage area in a circle of radius $R$, see Fig.~\ref{Fig1}.
The OWC AP is positioned at a height $L$ above the ground plane and the location of the $i$-th user relative to the AP is determined with the angle of irradiance $\theta_i$, the angle $\varphi_i$, and the radius $r_i$ in the polar coordinate plane.
We also introduce $\psi_i$ as the angle of incidence and denote the Euclidean distance between the LED lamp and the PD receiver by $d_i$. 

The DC channel gain of the LoS link between the $i$-th user and the PD receiver can be determined as \cite{IR}
\begin{equation}
h_i = \frac{ A_r\left( m + 1\right) R_r}{2\pi d_i^2}\cos ^m\left( \theta_i \right)T_sg \left( \psi_i \right)  \cos \left( \psi_i \right),
\label{I_n1}
\end{equation}
where $ A_r $ is the physical surface area of PD receiver, $R_r$ is the responsivity, and $T_s$ is the gain of the optical filter.
The optical concentrator is modeled as $g\left( \psi_i \right) =\zeta^2 /\sin^2\left( \Psi \right)$, for $0\leq\psi_i \leq \Psi$, where $\zeta$ is the refractive index of the lens at the PD and $\Psi$ denotes its field of view (FOV).
The LED transmission follows a generalized Lambertian radiation pattern with order $m = -\ln 2/ \ln \left( \cos \Phi_{1/2} \right)$, where $\Phi_{1/2}$ denotes the semi-angle at the half illuminance of LED\footnote{We assume that all LEDs are characterized by the same parameters, i.e., $m_i=m$ and $\Phi^i_{1/2}=\Phi_{1/2}$ for $\forall i$.}~\cite{OWC_MATLAB}.

Assuming that the surface of PD receiver is parallel to the ground plane where the IoT devices are located and is not oriented towards the LED, then $\theta_i =\psi_i $,   $d_i=~\sqrt {r_i^2 + L^2} $,    $\cos \left( \theta_i \right) \!=\! \frac{L}{ \sqrt {r_i^2 + L^2} }$, and (\ref{I_n1}) can be rewritten as
\begin{equation}
h_i = \frac{\mathcal X  }{\left( r_i^2 + L^2 \right)^{\frac{m + 3}{2}}}
\label{I_n2}
\end{equation}
where $\mathcal X = \frac{A_r\left( m + 1 \right)R_r}{2\pi}T_s g\left( \psi_i \right)L^{m + 1}$.

Given the uniform distribution of the IoT users within the circle of radius $R$, the PDF of the radial distance from the centre of the circle is \cite{SA3}
\begin{equation}
f_{r_i}\left( r \right) = \frac{2r}{R^2},\quad 0\leq r \leq R.
\label{pdf_rn}
\end{equation}
Using \eqref{I_n2} and \eqref{pdf_rn} and RV transformation techniques, the PDF of the channel gain, $h_i$, can be derived as
\begin{equation}
f_{h_i}\left( h \right) = \frac{2 \mathcal X ^{\frac{2}{m + 3}}}{R^2\left( m + 3 \right)}  h^{ - \frac{m + 5}{m + 3}},\quad  h_{\min }\leq  h \leq  h_{\max }
\label{pdf_In}
\end{equation}
where $ h_{\min} = \frac{\mathcal X }{{\left( R^2 + L^2 \right)}^{\frac{m + 3}{2}}}$ and  $h_{\max} = \frac{\mathcal X}{L^{m + 3}}$.

Similarly, the PDF of $\gamma_i$, defined in \eqref{gUa}, can be derived as~\cite{SA3}
\begin{equation}
f_{\gamma_i}\left( \gamma  \right) = \frac{(\mu\mathcal X^2)^{\frac{1}{m + 3}}}{R^2\left( m + 3 \right)}\gamma^{ - \frac{m + 4}{m + 3}},\quad \gamma_{\min }\leq \gamma \leq \gamma_{\max }
\label{pdf_hi2}
\end{equation}
where $\gamma_{\min} = \frac{\mu \mathcal X^2}{{\left(R^2 + L^2 \right)}^{m+ 3}}$,   $\gamma_{\max} = \frac{ \mu \mathcal X^2}{L^{2 \left( m+ 3\right)}}$, and $\mu= \frac{ P_t^2 \eta^2 }{\sigma _n^2}$.


The statistics of the overall SINR can be determined based on the characteristic function (CF) approach~\cite{Papoulis}. 
More precisely, the CF of $\gamma_i = \frac{ P_t^2 \eta^2  h_i^2}{\sigma _n^2}$ can be derived via (\ref{pdf_hi2}) as
\begin{equation}
\begin{split}
& \varphi_{\gamma_i} \left( t \right)  \triangleq  \E \left[ e^{jt\gamma_i} \right] = \int_{-\infty}^{\infty} e^{jt\gamma}  f_{\gamma_i}\left( \gamma  \right) \,\mathrm{d}\gamma \\
& = \frac{(\mu\mathcal X^2)^{\frac{1}{m + 3}}}{R^2\left( m + 3 \right)} \int_{\gamma_{\min}}^{\gamma_{\max}} \gamma^{ - \frac{m + 4}{m + 3}} e^{jt\gamma}   \,\mathrm{d}\gamma  = \frac{(\mu\mathcal X^2)^{\frac{1}{m + 3}}}{R^2\left( m + 3 \right)} \\
& \times\left( \Gamma\!\left(-\frac{1}{m+3},-j t \gamma_{\rm min}  \right) \!\! -\! \Gamma\! \left(-\frac{1}{m+3}, -j t \gamma_{\rm max}\!  \right) \!\! \right)
\end{split},
\label{cf_gi}
\end{equation}
where $\Gamma\left(\cdot,\cdot\right)$ is the upper incomplete gamma function  \cite[(8.35)]{grad}.

\subsection{Contribution to the SINR Statistics from the Reference User}

For the reference user (recall that it was adopted that its index is $i = U_a$), the PDF of $\gamma_{U_a}$ is determined in (\ref{pdf_hi2}) as
\begin{equation}
f_{\gamma_{U_a}}\left( \gamma  \right) = \frac{(\mu\mathcal X^2)^{\frac{1}{m + 3}}}{R^2\left( m + 3 \right)}\gamma^{ - \frac{m + 4}{m + 3}},\quad \gamma_{\min }\leq \gamma \leq \gamma_{\max }.
\label{pdf_gUa}
\end{equation}

\subsection{Contribution to the SINR Statistics from the Interfering Users} 

The channel gains $h_i$ (thus also $\gamma_i$) are assumed to be independent and identical distributed (i.i.d.) RVs, thus the CF of $\gamma_{\rm I} = \sum_{i=1}^{U_a-1} \gamma_i$ can be determined as~\cite{Papoulis}
\begin{equation}
\begin{split}
 \varphi_{\gamma_{\rm I}} \left( t \right) & \triangleq  \E \left[ e^{jt\gamma_{\rm I}} \right] = \E \left[ e^{jt \sum_{i=1}^{U_a-1} \gamma_i } \right] = \E \left[ \prod_{i=1}^{U_a-1} e^{jt \gamma_i  } \right] \\
& =  \prod_{i=1}^{U_a-1}  \E \left[ e^{jt \gamma_i  } \right]  =  \prod_{i=1}^{U_a-1} \varphi_{\gamma_i} \left( t \right) =   \varphi_{\gamma_i}^{U_a-1}  \left( t \right)
\end{split}
\label{cf_isi}
\end{equation}
where the CF of  $\gamma_i$ is defined in (\ref{cf_gi}).
The PDF of $\gamma_{\rm I}$ can be determined as
\begin{equation}
\begin{split}
 f_{\gamma_{\rm I}} \left( \gamma \right) &  = \frac{1}{2\pi} \int_{-\infty}^{\infty} e^{-jt\gamma}  \varphi_{\gamma_{\rm I}} \left( t \right) \,\mathrm{d}t  \\
& = \frac{1}{2\pi} \int_{-\infty}^{\infty} e^{-jt\gamma}  \varphi_{\gamma_i}^{U_a-1}  \left( t \right) \,\mathrm{d}t
\end{split}
\label{pdf_isi}
\end{equation}
for $\gamma_{\min }^{U_a-1}\leq \gamma \leq \gamma_{\max }^{U_a-1}$.

\subsection{Overall SINR Statistics of the Reference User}

First, we derive the PDF of the RV defined as $\lambda = \gamma_{\rm I} + 1$ as follows
\begin{equation}
\begin{split}
 f_{\lambda} \left( \lambda \right)  &  =  \frac{f_{\gamma_{\rm I}}\left(  \gamma_{\rm I} \right)}{\vert \frac{\mathrm{d}\lambda}{\mathrm{d} \gamma_{\rm I}} \vert } = f_{\gamma_{\rm I}} \left( \lambda -1 \right)  \\
 & = \frac{1}{2\pi} \int_{-\infty}^{\infty} e^{-jt\left( \lambda -1 \right)}  \varphi_{\gamma_i}^{U_a-1}  \left( t \right) \,\mathrm{d}t
\end{split}
\label{pdf_isi1}
\end{equation}
for $\gamma_{\min }^{U_a-1} + 1\leq \lambda \leq \gamma_{\max }^{U_a-1}+ 1$.  
Since {$ \gamma_{U_a} $ and $ \gamma_{\rm I} $ are independent  random variables,  the joint PDF is $f_{\gamma_{U_a},\lambda} \left( \gamma,\lambda \right)=f_{\gamma_{U_a}} \left( \gamma \right)f_{\lambda} \left( \lambda \right)$}.
The PDF of the ${\rm SINR} =\frac{\gamma_{U_a}}{\lambda}=\frac{\gamma_{U_a}}{\gamma_{\rm I} + 1}$ of the reference user, conditioned on the total number of active users $U_a$, can  be derived 
as follows~\cite{ratio}
\begin{equation}
\begin{split}
 f_{{\rm SINR} } \left( x | U_a \right) = \int_{-\infty}^{\infty}\vert\lambda \vert  f_{\gamma_{U_a}} \left(x\lambda \right) f_{\lambda} \left( \lambda \right) \mathrm{d}\lambda
\end{split},
\label{pdf_sinr}
\end{equation}
where $f_{\gamma_{U_a}} \left(\gamma \right)$ and $f_{\lambda } \left(\lambda  \right)$ are previously defined in (\ref{pdf_gUa}) and (\ref{pdf_isi1}), respectively. After replacing (\ref{pdf_gUa}) and (\ref{pdf_isi1}) in (\ref{pdf_sinr}), since $\lambda>0$, the PDF can be derived as
\begin{equation}
\begin{split}
 & f_{\rm SINR} \left( x | U_a \right) = \frac{(\mu\mathcal X^2)^{\frac{1}{m + 3}}}{2\pi R^2\left( m + 3 \right)} x^{ - \frac{m + 4}{m + 3}} \\
 & \!\times \int_{\gamma_{\min }^{U_a-1} + 1}^{\gamma_{\max }^{U_a-1} + 1}\!\!  \lambda ^{ - \frac{1}{m + 3}}  \left( \int_{-\infty}^{\infty} \!\! \!e^{-jt\left( \lambda -1 \right)}  \varphi_{\gamma_i}^{U_a-1}  \left( t \right) \,\mathrm{d}t \right) \mathrm{d}\lambda
\end{split}
\label{pdf_sinr1}
\end{equation}
where $\varphi_{\gamma_i}\left( t \right)$ is the CF previously defined in (\ref{cf_gi}).

The CDF of the ${\rm SINR}$ of the reference user, conditioned on $U_a$, can be derived as
\begin{equation}
\begin{split}
F_{{\rm SINR}}\left( \gamma  \right | U_a )= \int_{0}^{\gamma}  f_{\rm SINR } \left( x | U_a \right) \,\mathrm{d}x 
\end{split}.
\label{cdf_sinr}
\end{equation}


\section{Outage Probability: Numerical Results and Discussion}

In order to determine the system reliability in terms of the outage probability of a single transmission, we adopt the standard model that assumes the outage will happen if the overall SINR falls below a predetermined threshold $\gamma_{\rm th}$.
Based on the previously derived expressions, the outage probability of the transmission from a randomly selected user (i.e., the reference user), given that $U_a \geq 1$ users are active, can be calculated as
\begin{equation}
\begin{aligned}{ \rm P_{out}} ( U_a )  =\mathbb{P}\left[{ \rm SINR < \gamma_{\rm th}} |  U_a \right] = F_{{\rm SINR}}\left( \gamma_{\rm th}  | U_a \right) 
\end{aligned},
\label{Pout}
\end{equation}
where $ F_{\rm SINR } \left( \gamma_{\rm th} | U_a \right) $ is defined in (\ref{cdf_sinr}).
The unconditional outage probability is simply
\begin{align}
    {\rm P_{out}} = \sum_{n \geq 1} { \rm P_{out}} ( U_a ) \, \mathbb{P} [ U_a = n ].
\end{align}
For the demonstration purposes, we focus on a single slot and assume a Bernoulli arrival process, such that each of the $U$ users becomes activated in the slot with activation probability $p_a$, independently of any other user.
In this case
\begin{align}
    \mathbb{P} [ U_a = n ] = { U \choose n } p_a^n ( 1 - p_a)^{U-n}
\end{align}
i.e., $U_a$ is a binomial RV with mean $ U p_a$.

\begin{figure}[!b]
\centering
\includegraphics[width=\columnwidth]{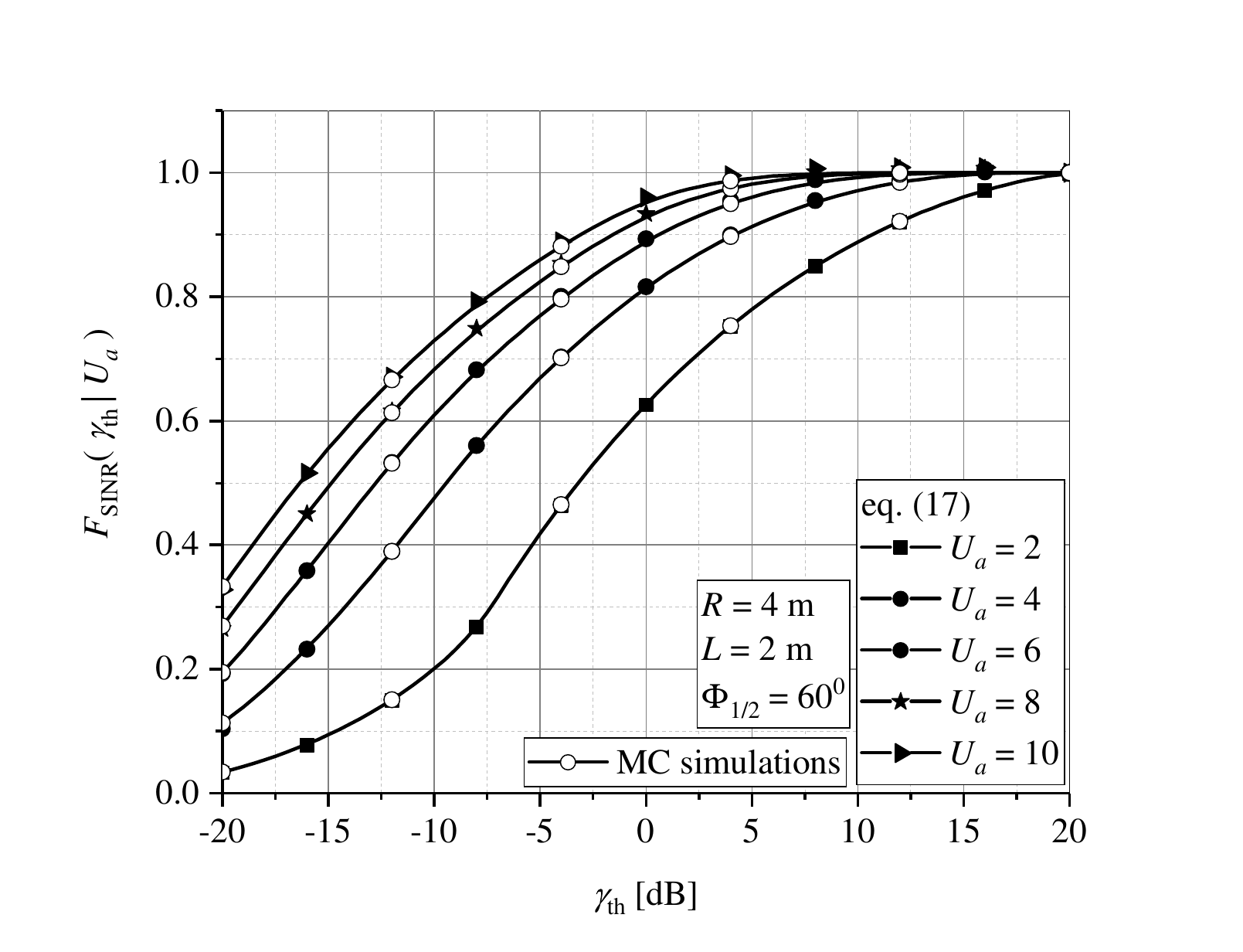}
\caption{CDF of the SINR of the reference user conditioned on $U_a$, $F_{{\rm SINR}}\left( \gamma_{\rm th} | U_a \right)$. }
\label{FigCDF}
\end{figure}

We evaluate this simple scenario by adopting the following values for the main parameters of the investigated setup: the FOV of the receiver is $90^{\circ}$, the PD surface area $A_r=1~{\rm cm}^2$, the responsivity $R_r=0.4~{\rm A}/{\rm W}$. The optical filter gain is $T_s =1$, and the refractive index of the lens at a PD is $\zeta =1.5$. The conversion efficiency is $\eta=0.8$, while the noise power spectral density takes a value $N_0=10^{-21}~{\rm W}/{\rm Hz}$, and the system bandwidth is chosen as $B=200~{\rm kHz}$. The transmitted optical power equals to $P_t = 30$ mW.

Note that both the PDF and CDF of the ${\rm SINR}$ in (\ref{pdf_sinr1}) and (\ref{cdf_sinr}), respectively, are given in integral form.
The present results are obtained by numerical computation of the derived expressions in MATLAB. 
The results are also confirmed through Monte Carlo (MC) simulations, as it can be observed in Fig.~\ref{FigCDF}, where the CDF of the ${\rm SINR}$ derived in (\ref{cdf_sinr}) is plotted.

\begin{figure}[!b]
\centering
\includegraphics[width=\columnwidth]{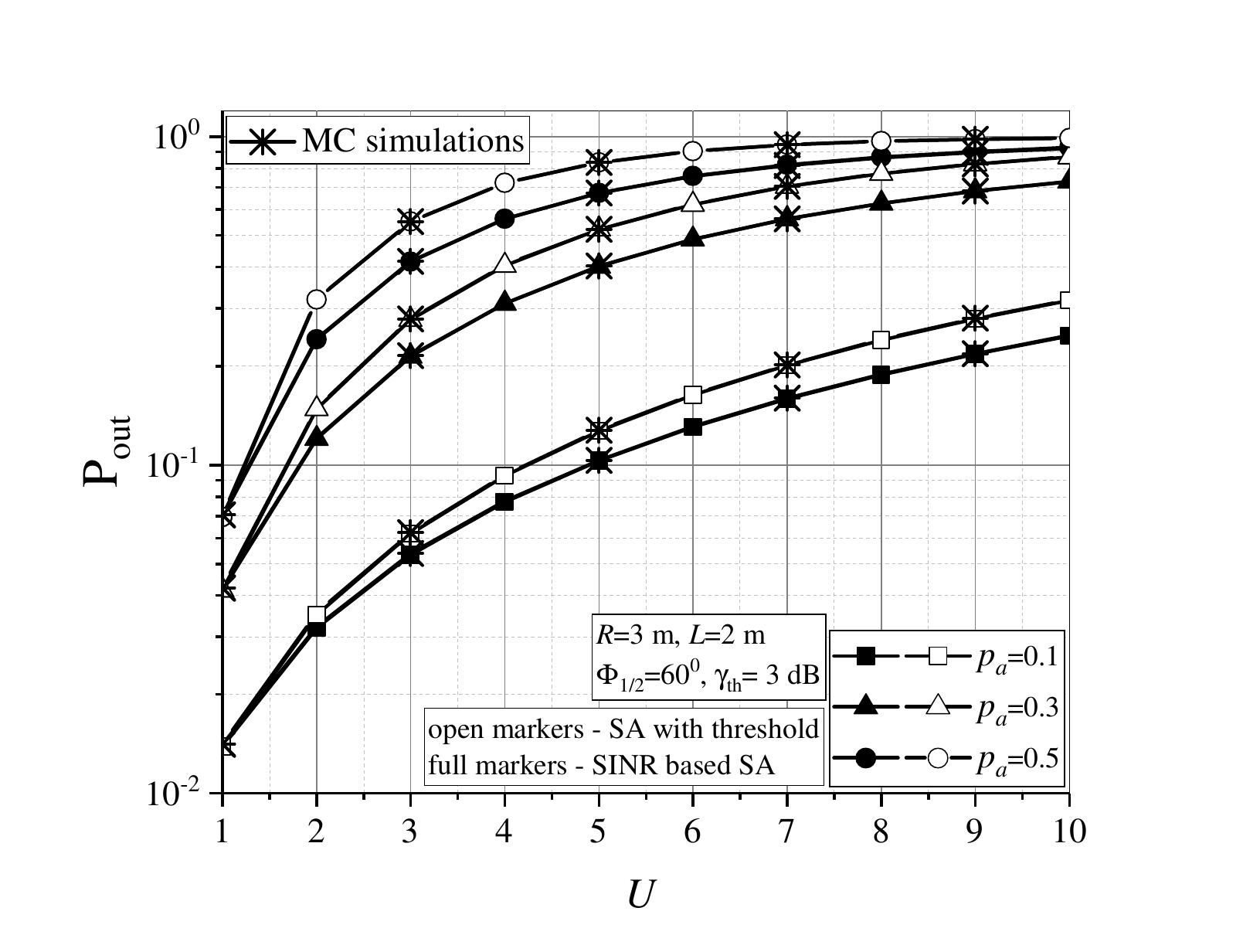}
\caption{Outage probability vs. total number of users for different values of  activation probability.}
\label{FigU}
\end{figure}

Fig.~\ref{FigU} depicts the outage probability of slot transmissions versus the total number of users, considering  three values of the activation probability $p_a$.
Besides the scenario observed in this paper, the classical SA scenario is also considered assuming that if more than one users are active, none of their packets will be decoded (i.e., collisions are by default destructive).
Additionally, if only the reference user is active, the received signal-to-noise ratio must be greater that the defined outage threshold $\gamma_{\rm th}$, i.e., the active user has to capture the channel at least against the noise.
The figure shows that the proposed system outperforms the classical scenario, i.e., the contribution of the instances when the reference user captures the collision slots is noteworthy. 
As it is expected, a greater value of $p_a$ determines a higher number of active users, thus the interference increases as well, and the packet has a lower chance to be successfully received.
With lower  $p_a$, a lower number of user is active, implying a higher probability of successful detection of the reference user.
We also note that the assumed value of the threshold ${\rm \gamma_{th}} = 3$~dB in Fig.~\ref{FigU} effectively implies that the power of the reference user has to be at least two times stronger than the combined power of the noise and the interference at the receiver, such that for lower values of ${\rm \gamma_{th}}$ (whenever it is enough for successful packet decoding) greater gains can be expected.

\begin{figure}[!t]
\centering
\includegraphics[width=\columnwidth]{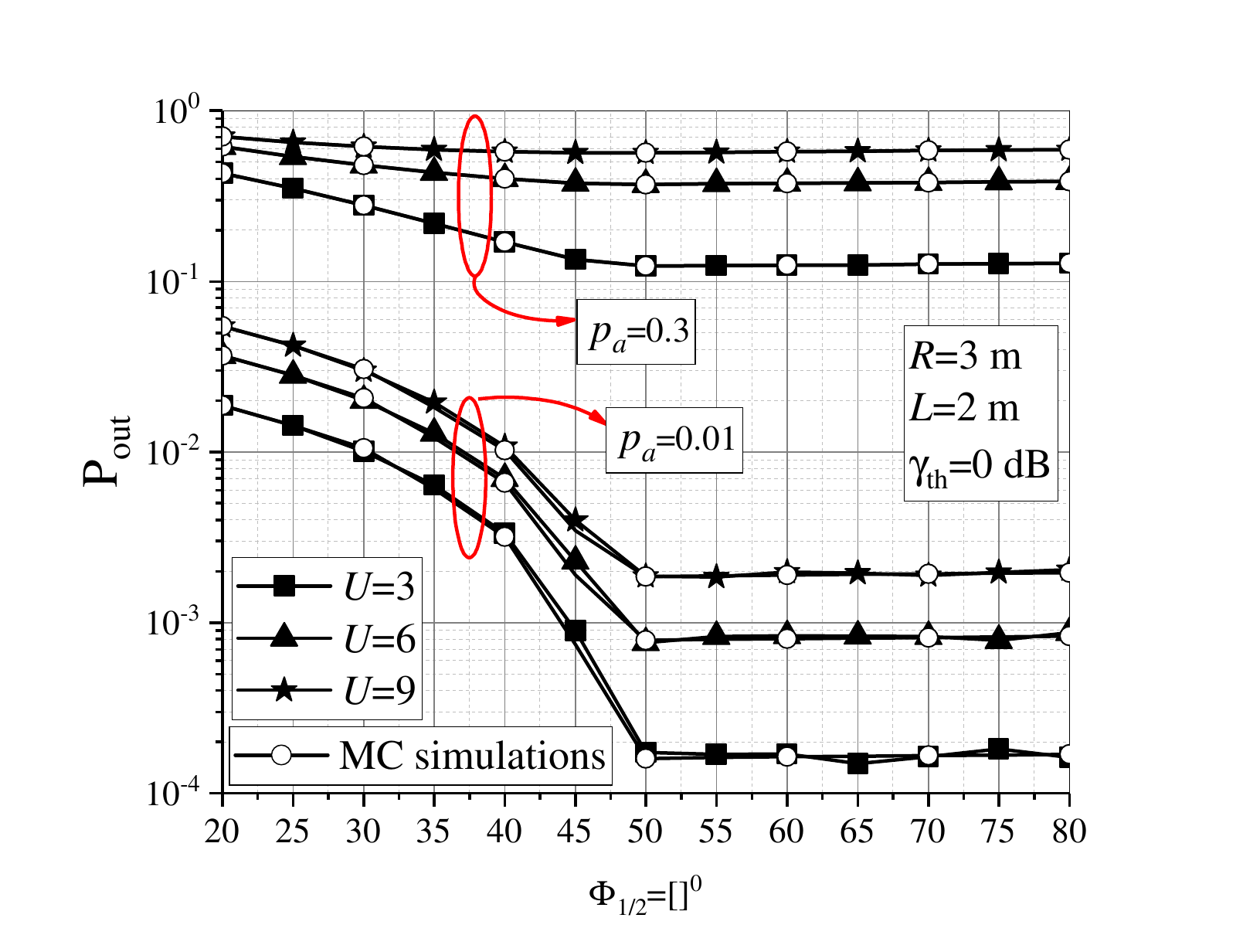}
\caption{Outage probability vs. semi-angle $\Phi_{1/2}$.}
\label{FigSA}
\end{figure}

Fig.~\ref{FigSA} shows the outage probability  dependence on the semi-angle at the half illuminance $\Phi_{1/2}$. Two values of activation probability are considered: $p_a=0.01$ and $p_a=0.3$. When the semi-angle takes a higher value, the optical beam at the LED output is wider, thus the signal strength at the AP from a distant user will be greater compared to the case when the semi-angle takes a lower value. This holds for all active users, so that it determines an impact on the interference contribution.  
When $p_a=0.01$, number of active users is very low, resulting in almost no interference, thus a high value for $\Phi_{1/2}$ will primarily have impact on the reference user, leading to an improved system performance. 
In the case of the higher activation probability, i.e., $p_a=0.3$, and higher number of active users $U$, the value of the semi-angle $\Phi_{1/2}$ has a slight impact on the outage probability, especially when the total number of users is very high.  
In that case, the interference contribution from other active users strongly affects the decoding process, resulting in a very low probability that the data  will be decoded.

The outage probability dependence on the radius $R$ is presented in Fig.~\ref{FigR}, assuming $p_a=0.01$ and $p_a=0.3$. The radius $R$ determines the size of the surface area where users are located. When the area where the IoT devices are placed is smaller, any user has a higher chance to be closer to the OWC AP, thus the received power at the PD will be greater compared to the case when the area is larger. When $p_a$ and $U$ take higher values, the number of active users may become very high, so that the interference contribution will result in a degraded system performance.
A higher $R$ value implies a very large surface area containing IoT devices, and therefore the reference user would probably be rather far from the OWC AP. This could make that the received power gets lower than the value required to meet the SINR threshold $\gamma_{\rm th}$, thus leading to a higher probability of outage.

\begin{figure}[!t]
\centering
\includegraphics[width=\columnwidth]{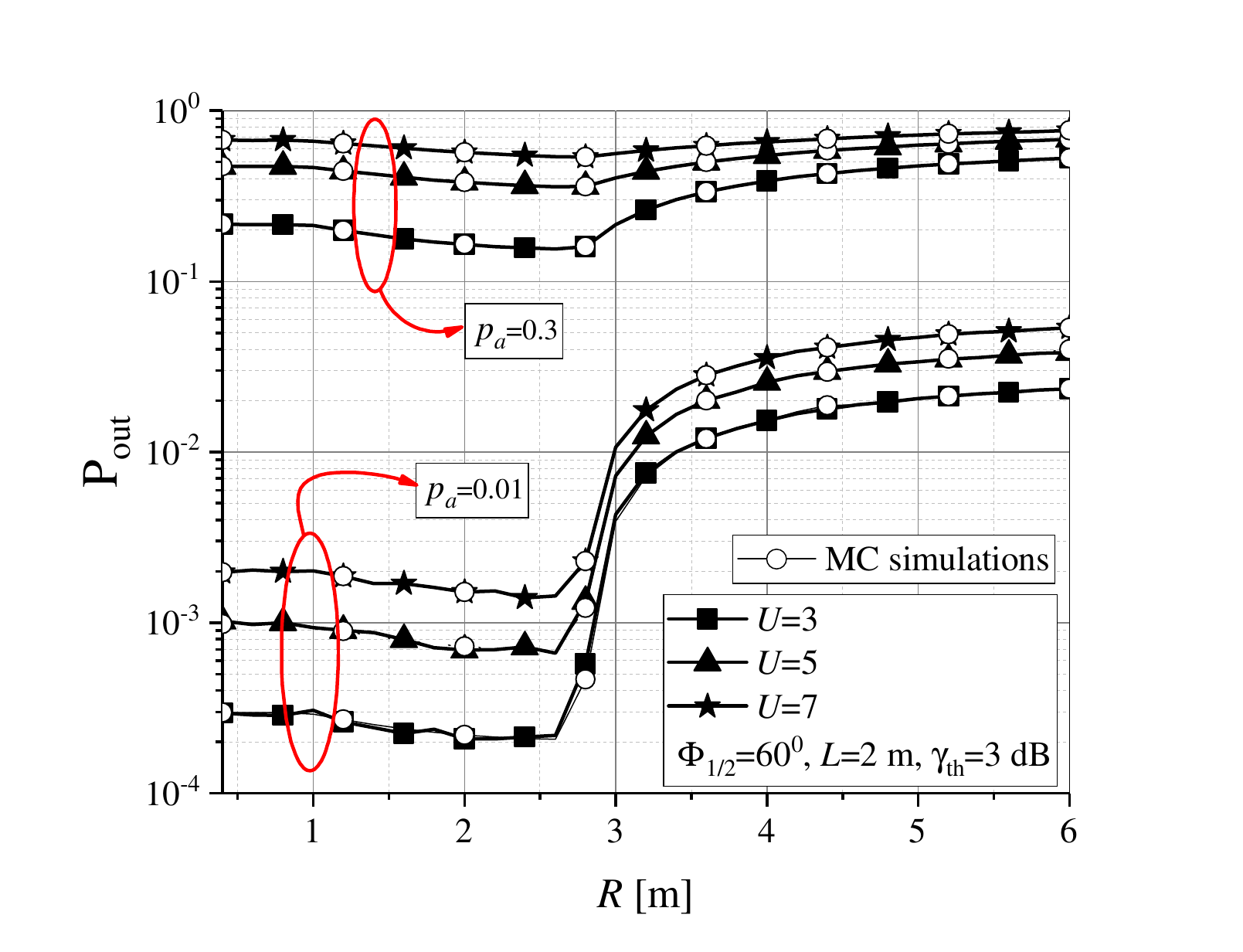}
\caption{Outage probability vs. radius $R$.}
\label{FigR}
\end{figure}

\section{Conclusion}

In this paper, we assess the reliability performance of a slotted ALOHA scheme for an indoor OWC-based IoT scenario.
By taking into account the effect of the collisions introduced by the interfering users contribution, the received SINR statistic has been derived, and we have been able to determine an analytical expression for the probability of outage of a transmission from an active user. Based on the derived expressions, meaningful numerical results have been presented and validated by MC simulation. The results show how the geometric setup of the IoT framework and the activation probability have an impact in the communication reliability of each user.
The analysis unfolded can be used to assess the performance of the random access protocol, which has been exemplified in the paper for the simple case of Bernoulli arrivals.
A significant potential application area of the current analysis is to use it for the optimization of the throughput performance of SA in OWC scenarios.
This topic is part of our further work.

\section*{Acknowledgment}

This work has received funding from the European Union Horizon 2020 research and innovation programme under the grant agreement No 856967. This work was supported by European Science
Foundation under COST  Action CA19111 (NEWFOCUS - European Network on Future Generation Optical Wireless Communication Technologies).

\end{document}